\documentclass[11pt]{article}
\usepackage{latexsym}  
\usepackage{amssymb}
\usepackage{graphicx}
\usepackage{amsmath}

\begin{document}

\hspace*{11cm} {MISC-2012-17 }

\begin{center}
{\Large\bf Large $\theta_{13}^\nu$ and Unified Description  }

{\Large\bf of Quark and Lepton Mixing Matrices}

\vspace{4mm}
{\bf Yoshio Koide$^a$ and Hiroyuki Nishiura$^b$}

${}^a$ {\it Department of Physics, Osaka University, 
Toyonaka, Osaka 560-0043, Japan} \\
{\it E-mail address: koide@het.phys.sci.osaka-u.ac.jp}

${}^b$ {\it Faculty of Information Science and Technology, 
Osaka Institute of Technology, 
Hirakata, Osaka 573-0196, Japan}\\
{\it E-mail address: nishiura@is.oit.ac.jp}

\date{\today}
\end{center}

\vspace{3mm}

\begin{abstract}
We  present a revised version of the so-called ``yukawaon model", which was proposed
for the purpose of a unified description of the lepton mixing matrix 
$U_{PMNS}$ and the quark mixing matrix $V_{CKM}$.  
It is assumed from a phenomenological 
point of view that the 
neutrino Dirac mass matrix $M_D$ is given
with a somewhat different structure from the charged lepton 
mass matrix $M_e$,  
although  $M_D=M_e$ was assumed in the previous model.  
As a result, the revised model predicts a reasonable value 
$\sin^2 2\theta_{13} \sim 0.07$ with keeping successful results for 
other parameters in $U_{PMNS}$ as well as $V_{CKM}$ and quark and lepton mass ratios. 
\end{abstract}

PCAC numbers:  
  11.30.Hv, 
  12.15.Ff, 
  14.60.Pq,  
  12.60.-i, 
\vspace{3mm}

\noindent{\large\bf 1 \ Introduction}

In a series of papers \cite{Yukawaon,K-N_EPJC12,K-N_PLB12}, 
the authors have investigated 
a unified description of 
the lepton mixing matrix \cite{PMNS} $U_{PMNS}$ and 
the quark mixing matrix \cite{CKM} $V_{CKM}$.
The essential idea is as follows: 
(i) The Yukawa coupling constants $Y_f$ ($f=u,d,e,$ and so on) 
in the standard model are effectively given by vacuum expectation 
values (VEVs) 
of scalars (``yukawaon") $Y_f$ with $3\times 3$ components, 
i.e. by $\langle Y_f \rangle /\Lambda$. Here $\Lambda$ is 
an energy scale of the effective theory. 
(The yukawaon model is a kind of the ``flavon" model \cite{flavon}.)
(ii) The model does not contain any coefficients which are 
dependent on the family numbers.  
The hierarchical structures of the effective Yukawa coupling 
constants originate only in a fundamental VEV matrix 
$\langle \Phi_0 \rangle$,
whose hierarchical structure is ad hoc assumed at present and 
whose VEV values are fixed by the observed charged lepton masses. 
(iii) Relations among those VEV matrices are obtained from 
SUSY vacuum conditions 
for a given superpotential under family symmetries and $R$ charges
assumed. (Since we use the observed charged lepton mass values
as the input values, it is a characteristic in the yukawaon model
that adjustable parameters are quite few.)

In the previous model\cite{Yukawaon,K-N_PLB12}, the quark and 
lepton mass matrices
(charged lepton mass matrix $M_e$, Dirac neutrino mass matrix 
$M_D$, down-quark mass matrix $M_d$, neutrino mass matrix $M_\nu$, 
and right-handed Majorana neutrino mass matrix $M_R$)  
are given as follows:
$$
\begin{array}{l}
M_e = k_e \Phi_0 ( {\bf 1} + a_e X_3 ) \Phi_0,  \\
M_D = M_e , \\
M_d = k_d \left[ \Phi_0 ( {\bf 1} + a_d X_3) \Phi_0 
+ m_d^0 {\bf 1} \right] ,  \\
M_u = k'_u \hat{M}_u \hat{M}_u , \\
\hat{M}_u = k_u \Phi_0 ( {\bf 1} + a_u X_3 ) \Phi_0,  \\
M_\nu = M_D M_R^{-1} M_D^T , \\
M_R = k_R (\hat{M}_u M_e + M_e \hat{M}_u ) + \cdots ,
\end{array}
\eqno(1.1)
$$
where $M_e$, $\Phi_0$, $X_3$, $\cdots$ are $3\times 3$ numerical matrices
which result from VEV matrices of scalar fields. 
Here the VEV matrices $\Phi_0$, $X_3$, and ${\bf 1}$ have structures given by
$$
\Phi_0 = \left( 
\begin{array}{ccc}
x_1 & 0 & 0 \\
0 & x_2 & 0 \\
0 & 0 & x_3 
\end{array} \right) , \ \ \ \ 
X_3 = \frac{1}{3} \left(
\begin{array}{ccc}
1 & 1 & 1 \\
1 & 1 & 1 \\
1 & 1 & 1 
\end{array} \right) , \ \ \ \ 
{\bf 1} = \left(
\begin{array}{ccc}
1 & 0 & 0 \\
0 & 1 & 0 \\
0 & 0 & 1 
\end{array} \right) .
\eqno(1.2)
$$
The coefficients $a_f$ ($f$ = e,u,d) which are important parameters in the model 
play an essential role in the mass ratios and mixings.  
On the other hand, the family-number independent coefficients $k_f$ and 
$k'_u$ do not any role in predicting family mixings and mass ratios. 
The values of $(x_1, x_2, x_3)$  with $x_1^2+x_2^2+x_3^2=1$ 
are fixed by the observed 
charged lepton mass values under the given value of $a_e$.
(In an earlier model \cite{O3_PLB09}, the charged lepton 
mass matrix $M_e$ was given by $M_e = k'_e \Phi_e \Phi_e$ 
and $M_d$ and $\hat{M}_u$ are given by those in (1.1) with the 
replacement $\Phi_0 \rightarrow \Phi_e$. 
The structures with $({\bf 1}+ a_f X_3)$ were suggested in  
a phenomenological model by Fusaoka and one of the authors
\cite{Democ_K-F_ZP96}.) 

The previous models \cite{Yukawaon,K-N_EPJC12,K-N_PLB12} have given almost successful 
unified description and predictions of $U_{PMNS}$ and $V_{CKM}$. 
However, these models have failed to give the observed large mixing of $\theta_{13}$ 
in $U_{PMNS}$: the observed value is $\sin^2 2 \theta_{13}
\sim 0.09$ \cite{theta13}, while the model in Ref.\cite{Yukawaon} predicts 
$\sin^2 2 \theta_{13} \sim 10^{-4}$. Even in a recent revised model \cite{K-N_PLB12}, 
the predicted value was, at most,  $\sin^2 2 \theta_{13} \sim 0.03$.
Since the model does not contain enough number of adjustable   
parameters as it is, it is hard to improve the prediction of 
$\sin^2 2 \theta_{13}$ without the cost of other successful predictions. 
So, an interesting attempt of introducing the structure $X_2$ into the 
model has been done in Ref.\cite{K-N_EPJC12}. 
In Ref.\cite{K-N_EPJC12}, the structure $X_2$ [see Eq.(1.44)] was 
introduced in $M_e$ together with assumption $M_D=M_e$,
but the predicted value of  $\sin^2 2 \theta_{13}$
was still small:  $\sin^2 2 \theta_{13} \sim 10^{-2}$. 
The $V_{CKM}$ was not discussed in Ref.\cite{K-N_EPJC12}.

In the present paper, we revise the model given in (1.1) by changing the structure only for the 
neutrino Dirac mass matrix $M_D$  
as follows; the structure $X_2$ is introduced in $M_D$ not in the charged lepton 
mass matrix $M_e$ unlikely in Ref.\cite{K-N_EPJC12}, and also 
it is assumed from a phenomenological 
point of view that the $M_D$ is given with a somewhat different coefficient from  $M_e$: 
$$
M_D = k_D \Phi_0 ( {\bf 1} + a_D X_2 ) \Phi_0,  
\eqno(1.3)
$$
where
$$
X_2 = \frac{1}{2} \left(
\begin{array}{ccc}
1 & 1 & 0 \\
1 & 1 & 0 \\
0 & 0 & 0 
\end{array} \right) . 
\eqno(1.4)
$$
Using this form we shall discuss $U_{PMNS}$ as well as $V_{CKM}$ of the model.
As to the structure $X_2$, we will discuss in Sec.2.
When once we  accept the form (1.3), we predict a reasonable value of
$\sin^2 2\theta_{13} \sim 0.07$ together with 
reasonable other parameters of $U_{PMNS}$, $V_{CKM}$
and quark and lepton mass ratios.

\vspace{3mm}

\noindent{\large\bf 2 \ Model}

We assume that a would-be Yukawa interaction is given as follows:
$$
W_Y = \frac{y_e}{\Lambda} e^c_i Y_e^{ij} \ell_j H_d 
+ \frac{y_\nu}{\Lambda} \nu^c_i Y_D^{ij} \ell_j H_u
+ \lambda_R \nu^c_i Y_R^{ij} \nu_j^c 
+ \frac{y_d}{\Lambda} d^c_i Y_d^{ij} q_j H_d  
+ \frac{y_u}{\Lambda} u^{c}_i Y_u^{ij} q_j H_u  ,
\eqno(2.1)
$$
where $\ell=(\nu_L, e_L)$ and $q=(u_L, d_L)$ are SU(2)$_L$ doublets.  
Under this definition of $Y_f$, the CKM mixing matrix and 
the PMNS mixing matrix are given by $V_{CKM}=U_u^\dagger U_d$ 
and $U_{PMNS}=U_e^\dagger U_\nu$, respectively, where
$U_f$ are defined by $U_f^\dagger M_f^\dagger M_f U_f
= D_f^2$ ($D_f$ are diagonal).  
(Hereafter, for simplicity, we denote $U_{PMNS}$ and $V_{CKM}$
as $U$ and $V$, respectively.)
In order to distinguish each yukawaon from others, we assume that 
$Y_f$ have different $R$ charges from each other together with $R$ 
charge conservation (a global U(1) symmetry in $N=1$ supersymmetry; 
for example, see Ref.\cite{SUSY}).
(Of course, the $R$ charge conservation is broken
at an energy scale $\Lambda'$.)

We assume the following superpotential for yukawaons 
by introducing fields $\Theta^f$, $P$, $E$, $\bar{E}$, $E'$, 
$\bar{E}^{\prime\prime}$, $E^{\prime\prime}$, 
$\bar{E}^{\prime}$, $\phi_e$, and $\phi_d$:  
$$
W_e= \lambda_e \left\{\phi_e Y_e^{ij} + \frac{1}{\Lambda} ({\Phi}_0)^{i\alpha}
 \left( (E^{\prime\prime})_{\alpha\beta} + a_e \frac{1}{\Lambda^2} 
X_{\alpha k} \bar{E}^{kl} X^T_{l\beta} \right) (\Phi_0^T)^{\beta j} 
\right\} \Theta^e_{ji} ,
\eqno(2.2)
$$
$$
W_D=  \frac{\lambda_D}{\Lambda} \left\{ (E')^\alpha_i Y_D^{ij} (E')^\beta_j 
+({\Phi}_0^T)^{\alpha i} \left(E_{ij} 
+ a_D \frac{1}{\Lambda^2} X^T_{j\gamma} (\bar{E}^{\prime\prime})^{\gamma\delta}
 X_{\delta j} \right)
(\Phi_0)^{j \beta} \right\} \Theta^D_{\beta\alpha} ,
\eqno(2.3)
$$
$$
W_u=  \frac{\lambda_u}{\Lambda} \left\{ {P}_{ik} Y_u^{kl} {P}_{lj} 
+ \hat{Y}^u_{ik} \bar{E}^{kl} \hat{Y}^u_{lj}\right\} \Theta_u^{ji},
\eqno(2.4)
$$
$$
W'_u = \frac{\lambda'_u}{\Lambda} 
\left\{ \bar{E}^{ik} \hat{Y}^u_{kl} \bar{E}^{lj}  +
({\Phi}_0)^{i\alpha} \left( (E^{\prime\prime})_{\alpha\beta} 
+ a_u \frac{1}{\Lambda^2}
X_{\alpha k} \bar{E}^{kl} X^T_{l\beta} \right) 
(\Phi_0^T)^{\beta j}\right\} \hat{\Theta}^{u}_{ji}, 
\eqno(2.5)
$$
$$
W_d= \lambda_d \left\{ \phi_d {Y}_d^{ij} 
+ \frac{1}{\Lambda} \left[
({\Phi}_0)^{i\alpha} \left( (E^{\prime\prime})_{\alpha\beta} 
+ a_d \frac{1}{\Lambda^2}
X_{\alpha k} \bar{E}^{kl} X^T_{l\beta} \right) (\Phi_0^T)^{\beta j} 
+ m_d^0 (\bar{E}')^i_\alpha (\bar{E}^{\prime\prime})^{\alpha\beta}
 (\bar{E}')^j_\beta
\right]
\right\} \Theta^{d}_{ji}, 
\eqno(2.6)
$$
$$
W_R= \left\{ \mu_R Y_R^{ij} + \frac{\lambda_R}{\Lambda} \left[ 
Y_e^{ik} \hat{Y}^u_{kl} \bar{E}^{lj} 
+ \bar{E}^{ik} \hat{Y}^u_{kl} Y_e^{lj} 
+ \xi_\nu^0 Y_D^{ik} E_{kl} Y_D^{lj} \right] \right\} \Theta^R_{ji} .
\eqno(2.7)
$$
Here, we have assumed family symmetries U(3)$\times$U(3)$'$. 
The fundamental yukawaon $\Phi_0$ is assigned to (3, 3) of U(3)$\times$ U(3)$'$, 
although quarks and leptons are still assigned to (3, 1) 
and yukawaons $Y_f$ are assigned to (6$^*$, 1) of U(3)$\times$ U(3)$'$.
In order to distinguish $R$ charges between $Y_e$ and $Y_d$, 
we have introduced U(3)$\times$U(3)$'$ singlet scalar fields
$\phi_e$ and $\phi_d$.

\begin{table}
\caption{Assignments of SU(2)$_L \times$SU(3)$_c \times$U(3)$
\times$U(3)$'$ and $R$ charges}

\vspace{2mm}
\begin{center}
\begin{tabular}{|c|cccccc|cc|} \hline
& $\ell$ & $e^c$ & $\nu^c$ & $q$ & $u^c$ & $d^c$ & 
$H_u$ & $H_d$  \\
\hline
SU(2)$_L$ & ${\bf 2}$ & ${\bf 1}$ & ${\bf 1}$ & ${\bf 2}$ & ${\bf 1}$ & 
${\bf 1}$ & ${\bf 2}$ & ${\bf 2}$  \\
SU(3)$_c$ & ${\bf 1}$ & ${\bf 1}$ & ${\bf 1}$ & ${\bf 3}$ & 
${\bf 3}^*$ & ${\bf 3}^*$ & ${\bf 1}$ & ${\bf 1}$  \\
U(3) & ${\bf 3}$ & ${\bf 3}$  & ${\bf 3}$ & ${\bf 3}$ & 
${\bf 3}$ & ${\bf 3}$ & ${\bf 1}$ & ${\bf 1}$   \\
U(3)$'$ & ${\bf 1}$ & ${\bf 1}$ & ${\bf 1}$ & ${\bf 1}$ & 
${\bf 1}$ & ${\bf 1}$ & ${\bf 1}$ & ${\bf 1}$   \\ 
$R$ & $r_\ell$ & $r_{ec}$ & $r_{\nu c}$ & $r_q$ & $r_{uc}$ & $r_{dc}$ & 
$r_{Hu}$ & $r_{Hd}$ \\ \hline
\end{tabular}
\begin{tabular}{|cccccc|cccccc|} \hline
$Y_e$ & $Y_D$ & $Y_R$ & $Y_u$ & $\hat{Y}^u$ & $Y_d$ & 
$\Theta^e$ & $\Theta^D$ & $\Theta^R$ & $\Theta_u$ & $\hat{\Theta}^{u}$ &
$\Theta^d$ \\ \hline
${\bf 1}$ & ${\bf 1}$ & ${\bf 1}$ & ${\bf 1}$ & ${\bf 1}$ & ${\bf 1}$ &
${\bf 1}$ & ${\bf 1}$ & ${\bf 1}$ & ${\bf 1}$ & ${\bf 1}$ & ${\bf 1}$ \\
${\bf 1}$ & ${\bf 1}$ & ${\bf 1}$ & ${\bf 1}$ & ${\bf 1}$ & ${\bf 1}$ &
${\bf 1}$ & ${\bf 1}$ & ${\bf 1}$ & ${\bf 1}$ & ${\bf 1}$ & ${\bf 1}$ \\
${\bf 6}^*$ & ${\bf 6}^*$ & ${\bf 6}^*$ & ${\bf 6}^*$ & ${\bf 6}^*$ & 
${\bf 6}^*$ & ${\bf 6}$ & ${\bf 1}$ & ${\bf 6}^*$ & ${\bf 6}$ & 
${\bf 6}$ & ${\bf 6}$ \\
${\bf 1}$ & ${\bf 1}$ & ${\bf 1}$ & ${\bf 1}$ & ${\bf 1}$ & ${\bf 1}$ &
${\bf 1}$ & ${\bf 6}$ & ${\bf 1}$ & ${\bf 1}$ & ${\bf 1}$ & ${\bf 1}$ \\
$r_{Ye}$ & $r_{YR}$ & $r_{YR}$ & $r_{Yu}$ & $\hat{r}_{Yu}$ & $r_{Yd}$ & 
$r_{\Theta e}$ & $r_{\Theta D}$ & $r_{\Theta R}$ & $r_{\Theta u}$ & 
$\hat{r}_{\Theta u}$ & $r_{\Theta d}$ 
\\ \hline
\end{tabular}
\begin{tabular}{|cc|cccccccc|cc|} \hline
$\Phi_0$ & $X$ & $E$ & $\bar{E}$ & $E'$ & $\bar{E}'$ & 
$E^{\prime\prime}$ & $\bar{E}^{\prime\prime}$ & $P$ & $\bar{P}$ & 
$\phi_e$ & $\phi_d$ \\ \hline
${\bf 1}$ & ${\bf 1}$ & ${\bf 1}$ & ${\bf 1}$ & ${\bf 1}$ & ${\bf 1}$ & 
${\bf 1}$ & ${\bf 1}$ & ${\bf 1}$ & ${\bf 1}$ & ${\bf 1}$ & ${\bf 1}$ \\
${\bf 1}$ & ${\bf 1}$ & ${\bf 1}$ & ${\bf 1}$ & ${\bf 1}$ & ${\bf 1}$ & 
${\bf 1}$ & ${\bf 1}$ & ${\bf 1}$ & ${\bf 1}$ & ${\bf 1}$ & ${\bf 1}$ \\
${\bf 3}^*$ & ${\bf 3}$ & ${\bf 6}$ & ${\bf 6}^*$ & ${\bf 3}$ & 
${\bf 3}^*$ & ${\bf 1}$ & ${\bf 1}$ & ${\bf 6}$ & ${\bf 6}^*$ &
${\bf 1}$ & ${\bf 1}$ \\
${\bf 3}^*$ & ${\bf 3}$ & ${\bf 6}$ & ${\bf 6}^*$ & ${\bf 3}$ & 
${\bf 3}^*$ & ${\bf 1}$ & ${\bf 1}$ & ${\bf 6}$ & ${\bf 6}^*$ &
${\bf 1}$ & ${\bf 1}$ \\
$r_0$ & $\frac{1}{2}(r_E+r^{\prime\prime}_E-1)$ & $r_{E}$ & $1-r_{E}$ &
 $r'_{E}$ & $1-r'_{E}$ & 
$r_{E}^{\prime\prime}$ & $1-r_{E}^{\prime\prime}$ & $r_{P}$ & $1-r_{P}$ &
$r_{\phi e}$ & $r_{\phi e}$ \\ \hline
\end{tabular}
\end{center}
\end{table}


We list the assignments of SU(2)$_L \times$SU(3)$_c \times$U(3)$
\times$U(3)$'$ and $R$ charges for the fields 
in the present model in Table 1. 
The assignments of $R$ charges are done so that the total $R$ charge 
of the superpotential term is $R(W)=2$.  
The $r$ parameters in Table 1 must satisfy the following relations:
 $r_{Hu}=2- r_\ell -r_D-r_{\nu c}-r_{Ye} =2-r_q - r_{uc} -r_{Yu}$, 
$r_{Hd}=2-r_\ell -r_{\nu c} -r_{Y d}=2-r_q -r_{uc} - r_{Yd}$, 
$r_{\Theta e}= 2-r_{Ye}-r_{\phi e}$, $r_{\Theta_D}=2-r_{YD}-2r'_E$,
$r_{\Theta_R}=2-r_{YR}$, $r_{\Theta_u}= 2-r_{Yu} -2r_P$, 
$\hat{r}_{\Theta_u}=1+r_E-\hat{r}_{Yu}$, and $r_{\Theta_d}=2-r_{Yd}-r_{\phi d}$.
Here, the $R$ charges of these fields must satisfy the following relations:
$2 r_0+r_E^{\prime\prime} = r_{Ye}+r_{\phi e} = r_{Yd}+r_{\phi d} =
\hat{r}_{Yu} +1 - r_E$, $2 r_0+r_E = r_{YD}+2r'_D$, and 
$r_{YR} = r_{Ye} +\hat{r}_{Yu} = 2 r_{YD} +r_E$. 
Since we consider that family symmetries U(3) and U(3)$'$ are 
gauge symmetries, the model must be anomaly free.
However, as seen in Table 1, the present model has anomaly coefficients
$A({\rm SU(3)})= 9$ and $A({\rm U(3)}') = 7$, so that 
we need further fields $({\bf 6}^* + {\bf 3}^* + {\bf 3}^*, {\bf 1})$ and
$( {\bf 1}, {\bf 6}^*)$ of U(3)$\times$U(3)$'$.
However, since roles of such additional fields in the 
present model are, at present, not clear, we do not 
discuss such fields.

From Eqs.(2.2) and (2.3) [and also (2.5) and (2.6)],
we obtain 
$$
R(E) +R(\bar{E}) = R(E^{\prime\prime}) + R(\bar{E}^{\prime\prime}).
\eqno(2.8)
$$
The VEVs of the introduced fields $E$, $\bar{E}$, $P$, and $\bar{P}$ are described by the
following superpotential by assuming $R(E\bar{E})=R(P\bar{P})=1$:
$$
W_{E,P} = \frac{\lambda_1}{\Lambda} {\rm Tr}[\bar{E} E \bar{P} P] 
+\frac{\lambda_2}{\Lambda} {\rm Tr}[\bar{E} E]  {\rm Tr}[\bar{P} P] ,
\eqno(2.9)
$$
which leads to 
$$
\langle E \rangle \langle \bar{E} \rangle \propto {\bf 1} , \ \ \ \ 
\langle P \rangle \langle \bar{P} \rangle \propto {\bf 1} .
\eqno(2.10)
$$
We assume specific solutions of Eq.(2.10):
$$
\frac{1}{v_E} \langle E \rangle = \frac{1}{\bar{v}_E} \langle \bar{E} \rangle 
= {\bf 1} , 
\eqno(2.11)
$$
$$
\frac{1}{v_P} \langle P \rangle = \frac{1}{\bar{v}_P^*} 
\langle \bar{P} \rangle^\dagger = {\rm diag}( e^{-i\phi_1}, e^{-i\phi_2}, 1),
\eqno(2.12)
$$
as the explicit forms of $\langle E \rangle$, $\langle \bar{E} \rangle$, 
and $\langle \bar{P} \rangle$.  
We assume similar superpotential forms for $E^{\prime\prime}$ and 
$\bar{E}^{\prime\prime}$, and for $E'$ and $\bar{E}'$. 

From SUSY vacuum conditions $\partial W/\partial \Theta =0$, 
we obtain the following relations:
$$
\langle Y_e \rangle = k_e \langle \Phi_0 \rangle \left( {\bf 1} + 
a_e X X^T \right) \langle \Phi_0^T \rangle ,
\eqno(2.13)
$$
$$
\langle Y_D \rangle = k_D \langle \Phi_0^T \rangle \left( {\bf 1} + 
a_D X^T X \right) \langle \Phi_0 \rangle ,
\eqno(2.14) 
$$
$$
\langle P \rangle \langle Y_u \rangle \langle P \rangle 
 = k'_u \langle \hat{Y}^u \rangle \langle \hat{Y}^u \rangle ,
 \eqno(2.15)
$$
$$
\langle \hat{Y}^u \rangle = k_u \langle \Phi_0 \rangle \left( {\bf 1} + 
a_u X X^T \right) \langle \Phi_0^T \rangle ,
\eqno(2.16)
$$
$$
\langle Y_d \rangle = k_d \left[ \langle \Phi_0 \rangle \left( {\bf 1} + 
a_d X X^T \right) \langle \Phi_0^T \rangle + m_d^0 {\bf 1} \right],
\eqno(2.17)
$$
$$
\langle Y_R \rangle = k_R \left( \langle Y_e \rangle 
\langle \hat{Y}^u \rangle + \langle \hat{Y}^u \rangle
\langle Y_e \rangle + \xi_\nu^0 
\langle Y_D \rangle \langle Y_D \rangle \right),
\eqno(2.18)
$$
where, for convenience, we have already put $\langle E \rangle$ 
as ${\bf 1}$, and so on. 
Here, since we have assumed that all $\Theta$ fields take 
$\langle\Theta\rangle =0$, we do not need to consider vacuum 
conditions for other fields $\partial W/\partial Y_e =0$, because
those always contain $\langle\Theta\rangle$. 
Thus, mass matrices are given by 
$M_e=\langle Y_e \rangle$, $M_D=\langle Y_D \rangle$, $M_u = \hat{k}_u \hat{M}_u
 \hat{M}_u$, $\hat{M}_u=\langle \hat{Y}^u \rangle$, $M_d=\langle Y_d \rangle$,
 $M_\nu = M_D M_R^{-1} M_D^T$, and $M_R=\langle Y_R \rangle$.

The most curious assumption is to assume the VEV matrix form of 
the scalar $X$ as 
$$
\frac{1}{v_X} \langle X \rangle_{\alpha i} = \frac{1}{2} \left(
\begin{array}{ccc}
1 & 1 & 0 \\
1 & 1 & 0 \\
1 & 1 & 0 
\end{array} \right)_{\alpha i} .
\eqno(2.19)
$$
The form (2.19) leads to
$$
\left(\langle X \rangle \langle X^T \rangle\right)_{\alpha\beta} = 
\frac{3}{2} (X_3)_{\alpha\beta}, \ \ \ \ 
\left(\langle X^T \rangle \langle X \rangle \right)_{ij} = 
\frac{3}{2} (X_2)_{ij}, 
\eqno(2.20)
$$
together with $\langle X \rangle \langle X \rangle=\langle X \rangle$, 
where $X_3$ and $X_2$ is defined by 
Eqs. (1.2) and (1.4), respectively, and, for simplicity, 
we have put $v_X=1$ because we are interested only in the relative 
ratios among the family components.      

At present, there is no idea for the origin of this form (2.19). 
We may speculate that this form is related to a breaking pattern
of U(3)$\times$U(3)$'$ (for example, 
discrete symmetries S$_2 \times$S$_3$). 
In the present paper, the form (2.12) is only ad hoc assumption. 
However, as seen later, we can obtain a good fitting for the neutrino 
mixing angle $\sin^2 2 \theta_{13}$ due to this assumption.

\vspace{3mm}

\noindent{\large\bf 3 \ Parameter fitting}

We summarize our mass matrices in the present model as follows:  
$$
M_e  = k_e \Phi_0 ( {\bf 1} + 
a_e X_3 )\Phi_0^T ,
\eqno(3.1)
$$
$$
M_D = k_D \Phi_0^T ( {\bf 1} + 
a_D X_2) \Phi_0 ,
\eqno(3.2) 
$$
$$
P M_u P = k'_u \hat{M}^u \hat{M}^u ,
 \eqno(3.3)
$$
$$
\hat{M}^u = k_u  \Phi_0  \left( {\bf 1} + 
a_u e^{i \alpha_u} X_3 \right) \Phi_0^T ,
\eqno(3.4)
$$
$$
M_d = k_d \left[  \Phi_0 ( {\bf 1} + a_d X_3) \Phi_0^T 
+ m_d^0 {\bf 1} \right],
\eqno(3.5)
$$
$$
M_\nu = M_D M_R^{-1} M_D^T,
\eqno(3.6)
$$
$$
M_R = k_R \left( M_e \hat{M}^u + \hat{M}^u M_e + \xi_\nu^0 
 M_D M_D \right),
\eqno(3.7)
$$
where, for convenience, we have dropped 
the notations ``$\langle$" and ``$\rangle$". 
Since we are interested only in the mass ratios and mixings, 
hereafter, we will use dimensionless expressions
$\Phi_0 = {\rm diag}(x_1, x_2, x_3)$,  
$P= {\rm diag} (e^{i\phi_1}, e^{i\phi_2},1)$, 
and $E={\rm diag}(1,1,1)$. 
For simplicity, we have regarded the parameter $a_d$ as
real correspondingly to the parameter $a_e$.
The parameters are re-refined by Eqs.(3.1)-(3.5). 

In the present model, we have 9 adjustable parameters 
except for $x_i$ 
[$a_e$, $a_D$, $(a_u, \alpha_u)$, $a_d$,
$(\phi_1, \phi_2)$, $m_d^0$, and $\xi_\nu^0$]  
for the 16 observable quantities (6 mass ratios in the
up-quark-, down-quark-, and neutrino-sectors, 4 CKM 
mixing parameters, and 4+2 PMNS mixing parameters). 
In order to fix these parameters, we use, as input values, 
the observed values for ${m_c}/{m_t}$, ${m_u}/{m_c}$, 
$\sin^2 2 \theta_{23}$, $\sin^2 2 \theta_{12}$, $R_\nu$, 
${m_d}/{m_s}$ as shown later. 
The relative ratios of parameters $x_i$ in $\Phi_0$ are fixed 
by the ratios of the charged lepton masses ${m_e}/{m_\mu}$ and
 ${m_\mu}/{m_\tau}$. 
The process of fixing parameters are summarized in Table. 2. 

\begin{table}
\caption{Process for fitting parameters. 
Of course, since these parameters listed in each step can slightly 
affect predicted values listed in the other steps, we need
fine tuning after the step 5th.  
}
\vspace{2mm}
\begin{center}
\begin{tabular}{|c|cc|cc|c|} \hline
Step & Inputs & $N_{inp}$ &  Parameters & $N_{par}$ &
 Predictions  \\ \hline
 &  $\frac{m_e}{m_\mu}$, $\frac{m_\mu}{m_\tau}$  
&  & $\frac{x_1}{x_2}$, $\frac{x_2}{x_3}$ & &  \\
1st  & $\frac{m_u}{m_c}$, $\frac{m_c}{m_t}$ & 5
& $a_e$, $a_u$ & 5 &    \\
    &  $\sin^2 2\theta_{23}$ &  & $\alpha_u$ & &  
\\ \hline
2nd  & $\sin^2 2\theta_{12}$  & 2 & $a_D$ & 2 &
$\sin^2 2\theta_{13}$, $\delta_{CP}^\ell$ , 2 Majorana phases \\
    &   $R_\nu$ &   & $\xi_\nu^0$  &  & $\frac{m_{\nu 1}}{m_{\nu 2}}$,
 $\frac{m_{\nu 2}}{m_{\nu 3}}$   \\ \hline
3rd  & $\frac{m_s}{m_b}$ & 1 & $a_d$ &  1 &   \\ \hline
4th  &  $|V_{us}|$, $|V_{cb}|$ & 2 & $(\phi_1, \phi_2)$ & 2 & 
$|V_{ub}|$, $|V_{td}|$, $\delta_{CP}^q$  \\ \hline
5th  & $\frac{m_d}{m_s}$ & 1 & $m_d^0$ & 1 &   
not affect to other predictions  \\ \hline 
option &  $\Delta m^2_{atm}$ &   & $m_{\nu 3}$ &  & 
$(m_{\nu 1}, m_{\nu 2},  m_{\nu 3})$, $\langle m \rangle$  \\
\hline
$\sum N_{\dots}$ & & 11 &  & 11 &   \\ \hline 
\end{tabular}
\end{center}
\end{table}

Now let us present the details of parameter fitting.    
Since we do not change the mass matrix structures 
for $M_e$, $M_u$, and $M_d$ from the previous paper
\cite{K-N_PLB12}, we use the following 
parameter values of $a_e$ and $(a_u, \alpha_u)$ 
$$
(a_e, a_u, \alpha_u) \sim (8, -1.35, \pm 8^\circ) ,
\eqno(3.8)
$$
which are fixed 
from the observed values of $m_c/m_t$, $m_u/m_c$, and
$\sin^2 2\theta_{23}$:
$$
r^u_{12} \equiv \sqrt{\frac{m_u}{m_c}} 
= 0.045^{+0.013}_{-0.010} , \ \ \ \ 
r^u_{23} \equiv \sqrt{\frac{m_c}{m_t}}
=0.060 \pm 0.005 ,
\eqno(3.9)
$$
at $\mu=m_Z$ \cite{q-mass}, and
$\sin^2 2\theta_{23} >0.95$ \cite{PDG12}. 
(These values will be fine-tuned in whole parameter
fitting of $U$ and $V$ later.)
Note that the neutrino sector of the model is different 
from the previous model, however 
the predicted values of $\sin^2 2\theta_{23}$ are almost 
the same before. 

\begin{figure}[h]
\begin{picture}(200,200)(0,0)

  \includegraphics[height=.3\textheight]{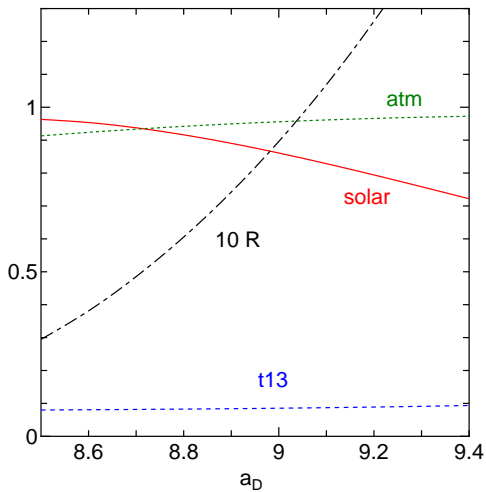}

\end{picture}  
  \caption{Lepton mixing parameters $\sin^2 2\theta_{12}$, 
$\sin^2 2\theta_{23}$, $\sin^2 2\theta_{13}$, and the neutrino 
mass squared ratio $R_\nu$ versus the parameter $a_D$. 
(``solar", ``atm", ``t13", and ``10 R" denote curves of 
$\sin^2 2\theta_{12}$, $\sin^2 2 \theta_{23}$, 
$\sin^2 2\theta_{13}$, and $R_\nu \times 10$, respectively.
Other parameter values are taken as 
$a_e=7.5$, $a_u=-1.35$, and $\alpha_u=7.6^\circ$. 
}
  \label{a_D}
\end{figure}

First, let us investigate lepton sector. 
In the revised model, a new parameter $a_D$ is 
added.
We illustrate the behaviors of Lepton mixing parameters 
$\sin^2 2\theta_{12}$, $\sin^2 2\theta_{23}$, 
$\sin^2 2\theta_{13}$, and the neutrino 
mass squared ratio $R_\nu$ versus the parameter $a_D$ 
for a case of $\xi_\nu^0=0$.
As seen in Fig.1, the parameter $a_D$ does not
change the prediction $\sin^2 2 \theta_{23} \sim 1$ 
in the previous model. 
Also, note that the prediction of $\sin^2 2 \theta_{13}$
is insensitive to the parameter $a_D$, i.e. 
$\sin^2 2 \theta_{13} \sim 0.08$. 
Only the predictions of $\sin^2 2\theta_{12}$ and 
$R_\nu = (m_{\nu2}^2 -m_{\nu1}^2)/(m_{\nu3}^2 -m_{\nu2}^2)$ are 
sensitive to the parameter $a_D$. 
In order to fit the observed value \cite{PDG12}
$\sin^2 2\theta_{12} = 0.857 \pm 0.024$,  
we take $a_D=9.01$.
However, in the model with $\xi_\nu^0=0$, 
the value $a_D=9.01$ cannot fit the 
observed value \cite{PDG12} of $R_\nu$, 
$$
R_{\nu} \equiv \frac{\Delta m_{21}^2}{\Delta m_{32}^2}=
\frac{(7.50\pm 0.20) \times 10^{-5}\ {\rm eV}^2}{
(2.32^{+0.12}_{-0.08}) \times 10^{-3}\ {\rm eV}^2} = 
(3.23^{+0.14}_{-0.19} ) \times 10^{-2} .
\eqno(3.10)
$$
The non-zero parameter $\xi_\nu^0$ has phenomenologically been brought 
in order to adjust the predicted value of $R_\nu$. 
The predicted values of $\sin^2 2 \theta_{23}$, 
$\sin^2 2\theta_{12}$, and $\sin^2 2 \theta_{13}$ are almost 
unchanged against the parameter $\xi_\nu^0$. 
In order to fit the neutrino mass ratio $R_\nu$, 
we take $\xi_\nu^0=-0.78$.

Next, we discuss quark sector. 
Since we have fixed the five parameters $a_e$, $a_u$, $\alpha_u$, 
$a_D$, and $\xi_\nu^0$, we have remaining four parameters for
six observables (2 down-quark mass ratios and 4 CKM mixing 
parameters). 
The parameters $a_d$ and $m_d^0$ are used to fit  
the observed down-quark mass ratios \cite{q-mass}
$$
r^d_{23} \equiv \frac{m_s}{m_b} = 0.019^{+0.006}_{-0.006} , \ \ \ 
r^d_{12} \equiv \frac{m_d}{m_s} = 0.053^{+0.005}_{-0.003} ,
\eqno(3.11)
$$
respectively.
Therefore, the four CKM mixing parameters are described  
only by two parameters $(\phi_1, \phi_2)$. 
As shown in Fig.~2, all the experimental constraints on 
CKM mixing matrix elements are satisfied by 
fine tuning with use of only two parameters $(\phi_1, \phi_2)$. 

\begin{figure}[t!]
  \includegraphics[width=70mm,clip]{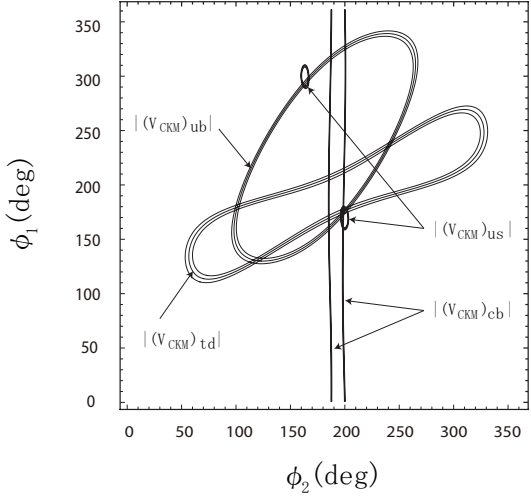}
  \caption{ Contour plots in the ($\phi_1$, $\phi_2$) parameter 
plane, which are shown by using experimental constraints on 
CKM mixing matrix elements:  
  $|V_{us}|=0.2252 \pm 0.0009$, $|V_{cb}|=0.0406\pm 0.0013$, 
$|V_{ub}|=0.00389 \pm 0.00044$, and  $|V_{td}|=0.0084 \pm 0.0006$.
The CKM elements depends on only the parameter set of 
[$a_e$, ($a_u$, $\alpha_u$), $a_d$, $m_d^0$, $\phi_1$, and $\phi_2$]. 
Here we present contour plots of the CKM elements 
in the ($\phi_1$, $\phi_2$) parameter 
plane by taking the values of other parameters as   
$a_e=7.5$, $a_u=-1.35$, $\alpha_u=-7.6^\circ$, $a_d=25$, and 
$m_d^0=0.0115$. 
We find that ($\phi_1$, $\phi_2$)=($177.0^\circ$, $197.4^\circ$) 
is consistent with all the CKM constraints.
}
  \label{Contour_Plots}
\end{figure}

Finally, we do fine-tuning of whole parameter values 
in order to give more improved fitting with the whole data. 
Our final result is as follows: 
under the parameter values
$$
a_e=7.5, \ a_D=9.01 , \ (a_u, \alpha_u) =(-1.35,-7.6^\circ), 
\  a_d=25, \ 
$$
$$
m_d^0=0.0115, \ \ \xi_\nu^0=-0.78,  \  (\phi_1,\phi_2)=(177.0^\circ,197.4^\circ), 
\eqno(3.12)
$$
we obtain 
$$
r^u_{12} =0.0465, \ \ \ r^u_{23}= 0.0614 , \ \ \ 
r^d_{12} = 0.0569, \ \ \ r^d_{23}=0.0240,
\eqno(3.13)
$$
$$
\sin^2 2\theta_{23}=0.969, \ \ \ 
\sin^2 2\theta_{12}= 0.860, \ \ \
\sin^2 2\theta_{13} = 0.0711 , \ \ \ 
R_\nu = 0.0324,
\eqno(3.14)
$$
$$
\delta_{CP}^{\ell}= -131^\circ  \ \ \ (J^{\ell} = -2.3 \times 10^{-2} ), 
\eqno(3.15)
$$
$$
|V_{us}|=0.2271, \ \ \ |V_{cb}|=0.0394, \ \ \ 
|V_{ub}|=0.00347, \ \ \ |V_{td}|= 0.00780 ,
\eqno(3.16)
$$
$$
\delta_{CP}^{q}= 59.6^\circ   \ \ \ (J^{q} = 2.6 \times 10^{-5} ). 
\eqno(3.17)
$$
Here, $\delta^\ell_{CP}$ and $\delta^q_{CP}$ are Dirac $CP$ 
violating phases in the standard conventions of $U$ 
and $V$, respectively.

Even if we choose a value of $\xi_\nu^0$ which gives a value of 
$R_\nu$ within $1\sigma$ given in Eq.(3.10), our 
predicted value of $\sin^2 2\theta_{13}$ is  
$\sin^2 2\theta_{13}=0.0711^{+0.003}_{-0.004}$, which 
is still somewhat small compared with the observed value 
$\sin^2 2\theta_{13} =0.098 \pm 0.013$ \cite{PDG12}.
So far, we have assumed that the parameter $\xi_\nu^0$ is real.
If we consider that the parameter $\xi_\nu^0$ is complex,
$\xi_\nu^0 \rightarrow \xi_\nu^0 e^{i \alpha_\nu}$, we can 
adjust the value of $\sin^2 2\theta_{13}$ without changing
other predicted values as seen in Fig.3. 
However, such a modification by the parameter $\alpha_\nu$ is 
not essential in the present model, 
so that we will regard that the parameter $\xi_\nu^0$ 
is real. 

\begin{figure}[h]
\begin{picture}(200,200)(0,0)

\includegraphics[height=.3\textheight]{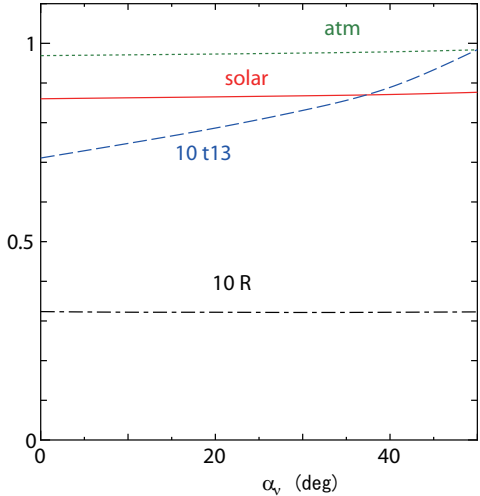}
\end{picture}  
  \caption{{Lepton mixing parameters $\sin^2 2\theta_{12}$, 
$\sin^2 2\theta_{23}$, $\sin^2 2\theta_{13}$, and the neutrino 
mass squared ratio $R_\nu$ versus the phase parameter $\alpha_\nu$. 
(``solar", ``atm", ``10 t13", and ``10 R" denote curves of 
$\sin^2 2\theta_{12}$, $\sin^2 2 \theta_{23}$, 
$\sin^2 2\theta_{13}\times 10$, and $R_\nu \times 10$, respectively.
Here the $\alpha_\nu$ dependence is presented under the parameter 
values given by (3.12).}
}
  \label{xi_0}
\end{figure}

We can also predict neutrino masses, for the parameters given (3.12) 
with real $\xi_\nu^0$, 
$$
m_{\nu 1} \simeq 0.0048\ {\rm eV}, \ \ m_{\nu 2} \simeq 0.0101 \ {\rm eV}, 
\ \ m_{\nu 3} \simeq 0.0503 \ {\rm eV}  ,
\eqno(3.18)
$$
by using the input value \cite{MINOS08}
$\Delta m^2_{32}\simeq 0.00243$ eV$^2$.
We also predict the effective Majorana neutrino mass \cite{Doi1981} 
$\langle m \rangle$ 
in the neutrinoless double beta decay as
$$
\langle m \rangle =\left|m_{\nu 1} (U_{e1})^2 +m_{\nu 2} 
(U_{e2})^2 +m_{\nu 3} (U_{e3})^2\right| 
\simeq 7.3 \times 10^{-4}\ {\rm eV}.
\eqno(3.19)
$$

Finally, let us comment on sensitivity of the predicted values Eq.(3.14)
to the input parameter values Eq.(3.12).
For simplicity, we show the sensitivity of only the lepton mixing and 
up-quark mass ratios to the input parameters $a_D$, $a_u$ and $\alpha_u$ 
in Table 3.
(We do not show sensitivity for the predicted CKM parameters,
because it can be easily seen in Fig.~3.)
In Table 3, values $\Delta x$ ($x=a_D$, $a_u$ and $\alpha_u$) 
for the parameter values $x$ are taken such as $(\Delta x)/x=0.05$, 
where the values $x$ are given in Eq.(3.12).
Here, we consider no change of values for the parameters $a_e$ and
$x_i$ ($i=1,2,3$)
because those have been fixed by the observed charged lepton 
masses with high accuracy.
We also do not discuss the parameter dependence of $R_\nu$ and 
$r^d_{12}=m_d/m_s$, because those are freely adjustable by the
parameters $\xi_\nu^0$ and $m_d^0$, respectively, without
almost affecting other observables. 
As seen in Table 3, the predicted value $\sin^2 2\theta_{13}$ is
sensitive to the parameter value $a_u$, so that we can take 
a value of $a_u$ which gives $\sin^2 2\theta_{13} \simeq 0.08$
at the cost of other fitting.
Also, we find that those predicted values are practically insensitive to 
the parameter value $\alpha_u$.  

\begin{table}
\caption{Sensitivity of the predicted values to the input parameter values. 
In the table, values $\Delta x$ ($x=a_D$, $a_u$ and $\alpha_u$) 
for the parameter values $x$
are taken such as $(\Delta x)/x=0.05$, where the values $x$ are
given in Eq.(3.12). 
Note that $r^u_{12}$ and $r^u_{23}$ are independent of the 
parameter $a_D$. 
}
\vspace{2mm}
\begin{center}
\begin{tabular}{|c|ccc|} \hline
 & $\Delta a_D= \pm 0.451$ & $\Delta a_u=\pm 0.068$ &  $\Delta \alpha_u=\pm 0.38$ \\ \hline
$r^u_{12}$ &  0.0465 & $ 0.0465^{+0.0239}_{-0.0179}$  & $ 0.0465^{+0.0023}_{-0.0022}$ \\
$r^u_{23}$ &  0.0614 & $ 0.0614^{+0.0075}_{-0.0054}$ & $ 0.0614^{+0.0017}_{-0.0016}$ \\
$\sin^2 2\theta_{12}$ & $ 0.860^{+0.092}_{-0.149}$ & $ 0.860^{+0.036}_{-0.045}$ & $ 0.860^{+0.004}_{-0.004}$ \\
$\sin^2 2\theta_{23}$ & $ 0.969^{+0.021}_{-0.040}$ & $ 0.969^{+0.023}_{-0.034}$ & $ 0.969^{+0.002}_{-0.002}$ \\
$\sin^2 2\theta_{13}$ & $ 0.0711^{+0.0012}_{-0.0016}$ & $ 0.0711^{+0.0094}_{-0.0091}$ & $ 0.0711^{+0.0001}_{-0.0001}$ \\
\hline 
\end{tabular}
\end{center}
\end{table}

\vspace{3mm}

\noindent{\large\bf 6 \ Concluding remarks}

In conclusion, by assuming VEV matrix forms of the yukawaons
Eqs.(3.1) -(3.7), we have obtained reasonable CKM and 
PMNS mixing parameters together with quark and neutrino 
mass ratios.
The major change from the previous yukawaon models is 
in the form of $M_D$. 
Although we give the form by assuming the VEV matrix $X_{\alpha i}$ which
is given by Eq.(2.19), and by considering the mechanism 
$(X X^T)_{\alpha\beta}=(X_3)_{\alpha\beta}$ versus 
$(X^T X)_{ij}=(X_2)_{ij}$, it is still phenomenological and 
somewhat factitious. 
However, when once we accept the form of $M_D$,
we can obtain $\sin^2 2\theta_{13} \sim 0.07$ whose value is 
not sensitive to the other parameters. 

Note that the present model does not have any family-dependent 
parameters except for $(x_1, x_2, x_3)$ in $\langle \Phi_0 \rangle$
and $(\phi_1, \phi_2)$ in $\langle P \rangle$. 
The parameter values $(x_1, x_2, x_3)$ have been fixed by 
the observed charged lepton masses.  
Therefore, the model have only 9 adjustable parameters for 16 observables. 
The 5 parameter values of 9 parameters, ($a_e$, $a_D$, $(a_u, \alpha_u)$, and 
$\xi_\nu^0$), have been fixed by the observed values $m_u/m_c$, $m_c/m_t$,
$\sin^2 2\theta_{12}$, $\sin^2 2\theta_{23}$, and $R_\nu$. 
Especially, the parameter $a_D$ has been fixed the observed value 
$\sin^2 2 \theta_{12}$. 
The parameter $\xi_\nu^0$ has been introduced only in order
to adjust the ratio $R_\nu$. 
(In other words, the value of $\xi_\nu^0$ has been fixed by
$R_\nu^{obs}$, the value (3.10).) 
Logically speaking, we need four observed values in order to
fix the four parameters $a_e$, $a_D$, and $(a_u, \alpha_u)$.
However, as seen in Fig.1, the values $\sin^2 2\theta_{23}\sim 0.9$
and $\sin^2 2\theta_{13} \sim 0.07$ are almost determined 
independently of the parameter $a_D$ when we fix $a_e$ 
and $(a_u, \alpha_u)$ from the observed up-quark mass ratios.
Therefore, $\sin^2 2\theta_{23}\sim 0.9$ and 
$\sin^2 2\theta_{13} \sim 0.07$ can substantially be regarded
as predictions in the present model. 
Of course, after we fix the 5 parameters, predictions are 
6 quantities: $\sin^2 2\theta_{13}$, 2 neutrino mass ratios, 
$CP$-violating phase parameter, and 2 Majorana phases.

Of the remaining 4 parameters $a_d$, $m_d^0$, and 
$(\phi_1, \phi_2)$, the parameters $a_d$ and $m_d^0$ 
are determined by the down-quark mass ratios 
$m_s/m_b$ and $m_d/m_s$, respectively.
Therefore, the 4 CKM mixing parameters are predicted 
only by adjusting the two parameters $(\phi_1, \phi_2)$.
We can obtain reasonable predictions of the CKM mixing parameters. 
 
The present model is still in a level of a phenomenological model. 
Nevertheless, it seems that 
the yukawaon model offers us 
a promising hint for a unified mass matrix model 
for quarks and leptons, i.e. it seems to suggest an idea 
that the observed family mixings and mass ratios of 
quarks and leptons are caused by a common origin.

\vspace{10mm}

{\Large\bf Acknowledgments}   

One of the authors (YK) would like to thank members of 
the particle physics group at NTU and NTSC, Taiwan, 
for the helpful discussions and their cordial hospitality.  
The work is supported by JSPS 
(No.\ 21540266).
%

\end{document}